\documentclass[
reprint,
aip, jap,
superscriptaddress,
floatfix
]{revtex4-1}

\usepackage{makeidx}
\usepackage[load=prefixed]{siunitx}
\usepackage{graphicx}
\usepackage{amsmath}
\usepackage{epstopdf}
\usepackage{textcomp}
\usepackage{epstopdf}
\usepackage{multirow}
\usepackage{array}
\usepackage{tabularx}
\usepackage{ulem} 
\usepackage{color}

\usepackage{natbib}
\usepackage{bm}%
\usepackage[colorlinks=true,linkcolor=blue]{hyperref}%
\expandafter\ifx\csname package@font\endcsname\relax\else
\expandafter\expandafter
\expandafter\usepackage
\expandafter\expandafter
\expandafter{\csname package@font\endcsname}%
\fi
\makeindex

\begin{document}

\title[Determination of the spin Hall angle in single-crystalline Pt films]{Determination of the spin Hall angle in single-crystalline Pt films from spin pumping experiments}

\address{Fachbereich Physik and Landesforschungszentrum OPTIMAS, Technische Universit\"{a}t Kaiserslautern,
	Erwin-Schr\"{o}dinger-Str. 56, 67663 Kaiserslautern, Germany}
\address{Physics Department, Aristotle University of Thessaloniki, 54124 Thessaloniki, Greece}

\author{Sascha Keller}
\author{Laura Mihalceanu}
\author{Matthias R. Schweizer}
\author{Philipp Lang}
\author{Bj{\"o}rn Heinz}
\author{Moritz Geilen}
\author{Thomas Br{\"a}cher}
\author{Philipp Pirro}
\author{Thomas Meyer}
\author{Andres Conca}

\address{Fachbereich Physik and Landesforschungszentrum OPTIMAS, Technische Universit\"{a}t Kaiserslautern,
	Erwin-Schr\"{o}dinger-Str. 56, 67663 Kaiserslautern, Germany}


\author{Dimitrios Karfaridis}
\author{Georgios Vourlias}
\author{Thomas Kehagias}

\address{Physics Department, Aristotle University of Thessaloniki, 54124 Thessaloniki, Greece}
%

\author{Burkard Hillebrands}

\author{Evangelos Th. Papaioannou}
\email{papaio@rhrk.uni-kl.de}
\address{Fachbereich Physik and Landesforschungszentrum OPTIMAS, Technische Universit\"{a}t Kaiserslautern,
	Erwin-Schr\"{o}dinger-Str. 56, 67663 Kaiserslautern, Germany}

\date{\today}

\begin{abstract}

\noindent We report on the determination of the spin Hall angle in ultra-clean, defect-reduced epitaxial Pt films. By applying vector network analyzer ferromagnetic resonance spectroscopy to a series of single crystalline Fe (12 nm) /Pt (t$_\text{Pt}$) bilayers we determine the real part of the spin mixing conductance $(4.4\:\pm\:0.2)\cdot 10^{19}\: $m$^{-2}$ and reveal a very small spin diffusion length in the epitaxial Pt $(1.1\:\pm\:0.1)$ nm film. We investigate the spin pumping and ISHE in a stripe microstucture excited by a microwave coplanar waveguide (CPW) antenna. By using their different angular dependencies, we distinguish between spin rectification effects and the inverse spin Hall effect. The relatively large value of the spin Hall angle $(5.7 \pm 1.4)\: \% $ shows that ultra-clean e-beam evaporated non-magnetic materials can also have a comparable spin-to-charge current conversion efficiency as sputtered high resistivity layers.

\end{abstract}



\maketitle {}

\section{Introduction}

In spintronics, bilayers composed of a ferromagnetic (FM) and a nonmagnetic (NM) layer with large spin-orbit-interaction are used to investigate spin-to-charge current conversion. At ferromagnetic resonance (FMR) the spin pumping (SP) effect results in the injection of a pure spin current into the NM layer \cite{Bauer2002} and the inverse spin Hall effect (ISHE) generates a charge current by spin-dependent deflection \cite{Saitoh2006}. For magnon spintronics the spin-to-charge current conversion may play an important role for its success since it can define the interface to common CMOS technology. Many aspects of the spin pumping effect using ferromagnetic metals \cite{Andres} and insulators \cite{goennewein2011} with Pt capping or Py with varying non-magnetic layers \cite{Mosendz2010, MosendzPRL2010} as well as metallic~\cite{Caminale2016} and dielectric interlayer~\cite{Mihalceanu2017} materials have been investigated.

\noindent The parameter describing the efficiency of the spin-to-charge current conversion is the spin Hall angle $\Theta_\text{SH}$. It is defined as the ratio of the measured charge current deflected by ISHE in the NM material and the spin current injected into the NM material. 
Another parameter, strongly correlated to the spin Hall angle, is the spin diffusion length $\lambda_\text{SD}$. $\lambda_\text{SD}$ describes the typical length scale in which the spin current is attenuated after its injection from the FM into the NM material. This attenuation in the NM is originates from spin-flip processes and from the deflection of the spin current caused by the ISHE.
For the commonly used NM material Pt, $\Theta_\text{SH}$ and $\lambda_\text{SD}$ have a wide span in the literature, namely from 1\%~up to 11 \% and from under 1 nm to roughly 10 nm, respectively~\cite{goennewein2011, Mosendz2010, MosendzPRL2010, Hoffmann2013, Rojas2014, azevedo, Soh2014, Wang2014, Nakayama2012, Feng2012, Vlaminck2013}. 
If the depolarization of the spin current by spin memory loss can be neglected, both parameters, $\Theta_\text{SH}$ and $\lambda_\text{SD}$, can show a reciprocal behavior~\cite{Rojas2014}: A small $\lambda_\text{SD}$ is then connected to a large $\Theta_\text{SH}$ and vice versa. Often, this correlation can be seen in the literature.
The experimental finding of $\lambda_\text{SD}$ is crucial for determining $\Theta_\text{SH}$  and will be discussed in this paper.

\noindent When a metallic ferromagnet is used, spin rectification (SR) effects, in particular anisotropic magnetoresistance and anomalous Hall effect, generate a DC voltage in the FM material at FMR. 
This DC voltage is added to the ISHE voltage making it difficult to disentangle the signals. There are various experimental methods to separate ISHE from SR effects, e.g. by applying a microwave cavity to minimize the microwave currents or making use of the external magnetic field dependence of the effects \cite{Saitoh2016, Harder2016}. In a recent publication we demonstrate the complexity of the results of an angle-resolved spin pumping setup with macroscopic bilayer samples \cite{Keller2017}. 

\noindent Most of the studies on metallic systems are devoted to sputtered polycrystalline samples, thus the influence of the crystal structure of those materials on the spin pumping experiments is neglected. Recently, studies on epitaxial bilayer systems are arising in literature~\cite{Andres, Keller2017, Huo2017}. In a prior publication~\cite{Evangelos}, we optimized the growth of epitaxial Fe/Pt bilayers for spin pumping experiments: We have shown that the real part of the spin mixing conductance $g^{\uparrow \downarrow}$ (parameter which quantifies the transport of angular momentum through the FM/NM interface) and the inverse spin Hall efficiency (inverse spin Hall current divided by the absorbed microwave power) could be increased by increasing annealing temperature from room temperature to 300$^\circ$C. With in-situ scanning tunneling microscope imaging it was shown, that with increasing annealing temperature the Fe/Pt interface roughness was also increased. With the present work, we aim to determine the spin Hall angle of epitaxial Fe/Pt at optimal annealing temperature (300$^\circ$C). 

\noindent In the first section, we focus on the growth of the Fe/Pt bilayers, via X-ray diffraction (XRD), high resolution transmission electron microscopy (HRTEM) imaging and selected area electron diffraction (SAED) analysis, and explain the microstructuring process and design. In the second part, we determine the spin diffusion length of the Pt and the real part of the spin mixing conductance of the Fe/Pt interface by vector network analyzer ferromagnetic resonance spectroscopy (VNA-FMR) measurements of unstructured Fe/Pt films. For this, we investigated the Gilbert damping parameter of the films with varying Pt thickness.  
In the third section, we show the results of the angle-resolved spin pumping measurements of the microstructured Fe(12nm)/Pt(6nm) sample. There, we discuss, how the ISHE and AMR voltages change, when the magnetization rotates in-plane, and, how the strong anisotropy of Fe influences the resonance condition and the ellipticity of the precession. In the last part, we calculate and discuss the spin Hall angle and compare the efficiency of the spin-to-charge current conversion in the epitaxial Pt  to the results provided by recent publications.

\section{Structural quality and microstructuring process of the Fe/Pt samples}

\begin{figure}
\begin{center}
 \includegraphics[width =0.9 \columnwidth]{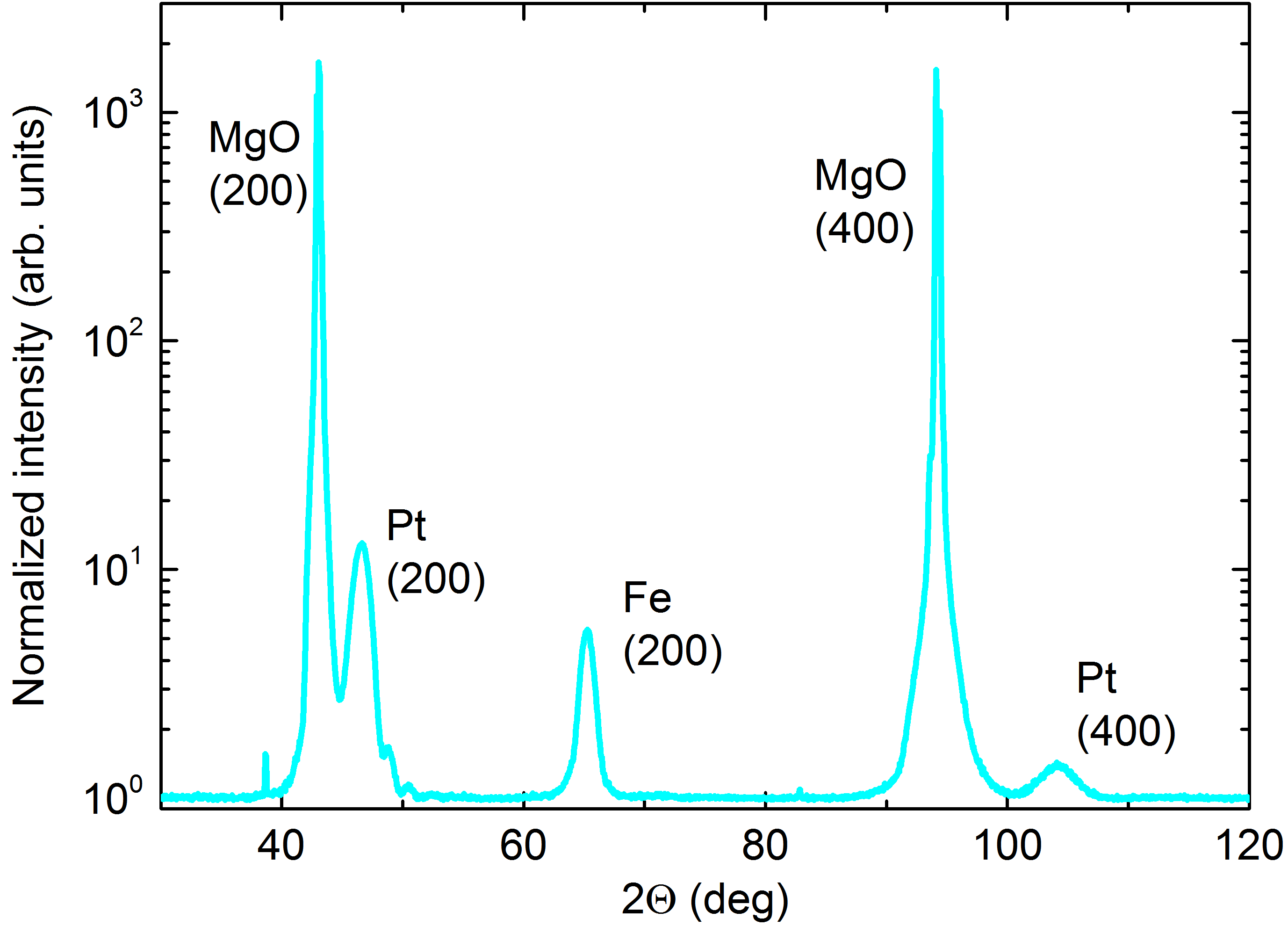}
\caption{\label{fig:xrd} X-ray diffraction (XRD) pattern of the unstructured Fe/Pt bilayer. It displays Pt (200) and (400) as well as Fe(200) diffraction peaks beside the MgO (200) and (400) peaks of the substrate showing the formation of single crystalline grains of both materials with distinct orientation.}
\end{center}
\end{figure}

\begin{figure}
\begin{center}
 \includegraphics[width =0.9 \columnwidth]{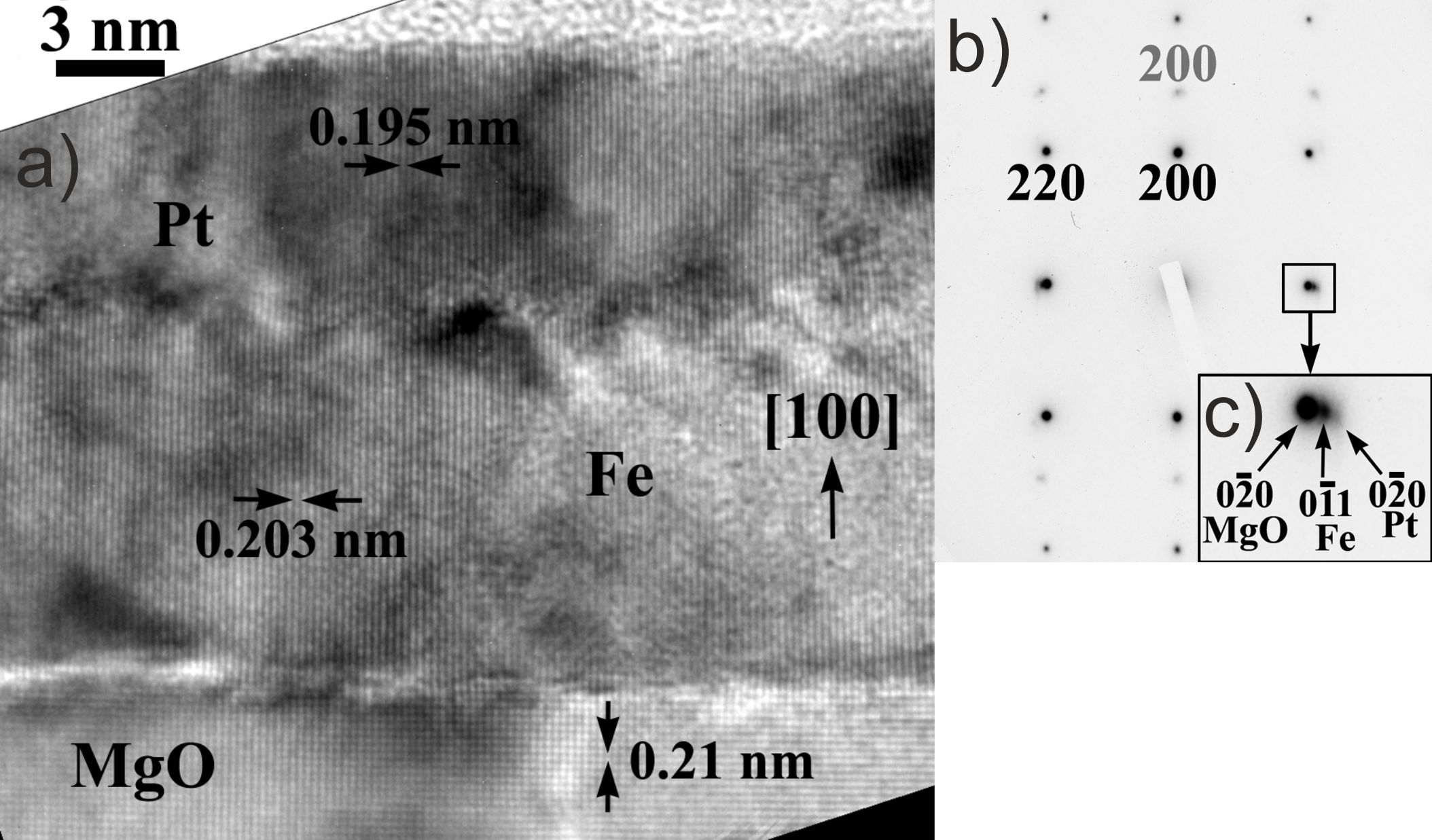}
\caption{\label{fig:tem} a) Cross-sectional HRTEM image of the Fe(12nm/Pt(6nm) bilayer on MgO(100), along the [001] projection direction. Perfect alignment of the in-plane crystal planes in all lattices is evident, while their experimental \textit{d}-spacing values are close to their stress-free bulk values. b) Corresponding common SAED pattern showing the good alignment of the (200) MgO and (200) Fe (grey indices) crystal planes along the growth direction. c) In the magnified part, the perfect epitaxy of the ($0\overline{2}0$) MgO, ($0\overline{1}1$) Fe and ($0\overline{2}0$) Pt planes (in-plane direction) is also shown. Occasionally, the Pt lattice slightly deviates towards the [110] direction.}
\end{center}
\end{figure}

Fe/Pt bilayers with a constant Fe thickness of 12 nm and a varying Pt thickness have been grown by molecular beam epitaxy (MBE) on MgO (100) substrates with a growth temperature of 300 C$^\circ$ in an ultra-high vacuum system with a base pressure of $5 \cdot 10^{-10}\: $mBar. To prevent oxidization we additionally capped the samples with Pt thicknesses below $1$ nm with a $10\: $nm MgO layer. We have previously shown that this MgO capping layer has no influence on the Gilbert damping parameter $\alpha$ in ferromagnetic resonance spectroscopy measurements \cite{Andres}. To confirm the epitaxial growth of the samples we performed XRD and HRTEM measurements. The XRD pattern of the Fe(12 nm)/Pt(6 nm) sample is indicatively shown in Fig.~\ref{fig:xrd}: Pt (200) and (400) as well as Fe (200) diffraction peaks beside the MgO (200) and (400) peaks of the substrate are present, showing the formation of single crystalline grains of both materials with distinct orientation. In Fig.~\ref{fig:tem}a), a HRTEM image of the Fe(12nm)/Pt(6nm) sample is depicted, illustrating the epitaxial structural quality of the individual layers that exhibit a single crystalline nature. 
The Fe/Pt interface is coherent and continuous across the bilayer presenting an \textit{rms} roughness of (0.7 $\pm$ 0.1) nm (2 to 3 Fe monolayers), while the Pt layer compensates this roughness, exhibiting a minimal \textit{rms} surface roughness of (0.2 $\pm$ 0.02) nm. Later we discuss, how the atomic roughness of the Fe/Pt interface influences the ISHE in epitaxial Pt. The HRTEM imaging along with SAED analysis (Fig.~\ref{fig:tem}b) showed the perfect epitaxial growth relation of the bilayer on the MgO substrate, considering that the Fe lattice is 45$^\circ$ in-plane rotated relative to the MgO and Pt lattices. Moreover, HRTEM measurements demonstrated that the resolved interplanar \textit{d}-spacing values of the involved lattices are close to their bulk values and, therefore, a relaxed structural configuration is anticipated. This implies that the mismatch between consecutive lattices is accommodated by misfit dislocations.

\begin{figure}
\begin{center}
\includegraphics[width =0.9 \columnwidth]{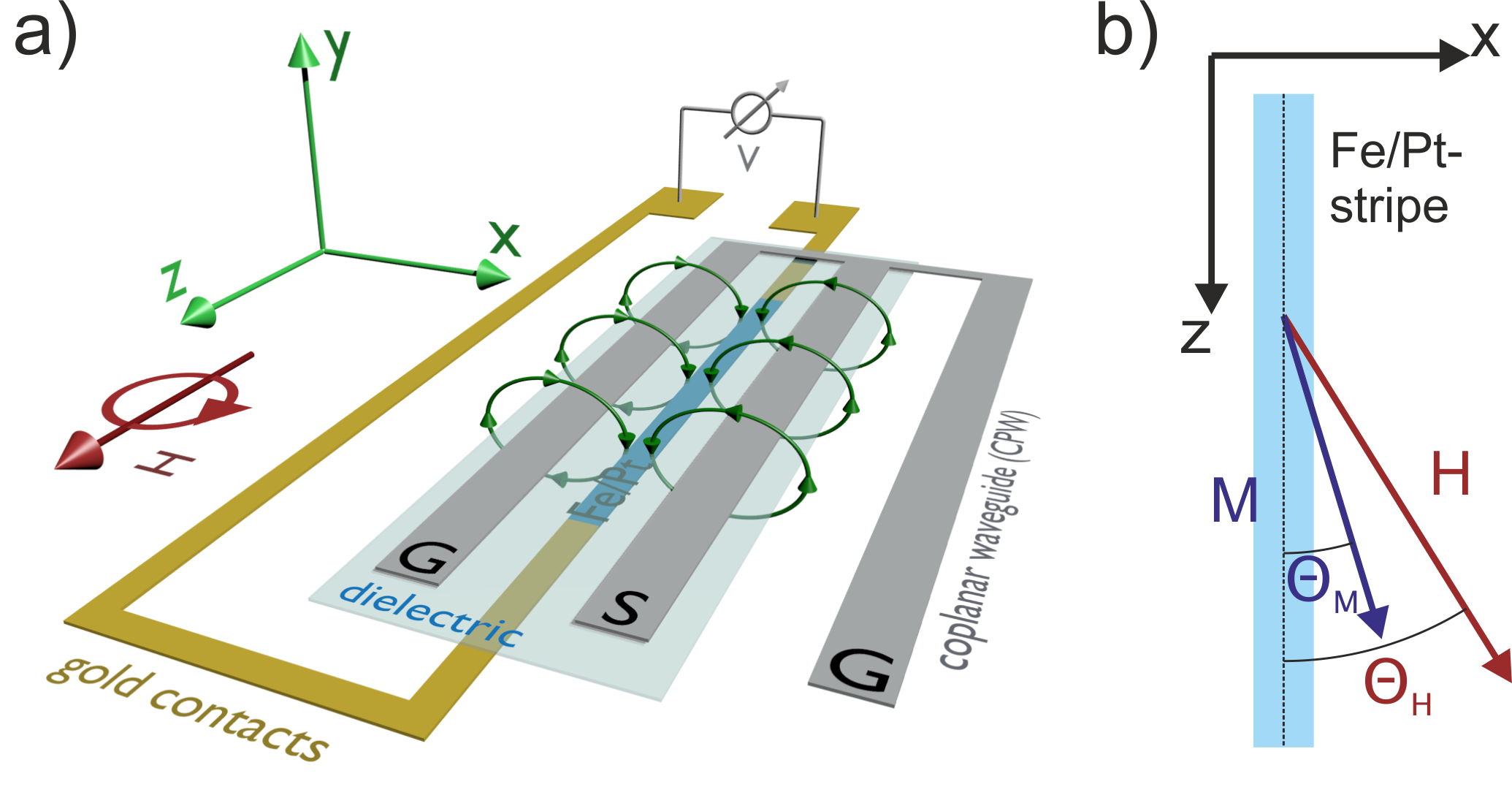}
\caption{\label{fig:oopCPW}a) Scheme of the microstructured Fe(12nm)/Pt(6nm) sample: The plane film was structured into 200 \textmu m$\times$10 \textmu m stripes along the Fe (010) axis.   Afterwards, Cr/Au/Cr contacts for the Fe/Pt stripes, a dielectric layer (SiO$_2$) and a coplanar waveguide(CPW)-like Ti/Cu/Ti antenna structure have been fabricated. The dielectric layer of 220 nm prevents DC currents flowing between the Cr/Au/Cr contacts and the microwave antenna. The CPW antenna provides an out-of-plane Oersted field, which excites the FMR inside the Fe film. Caused by the spin pumping effect and ISHE, as well as by spin rectification effects a DC voltage arises, which can be measured by the Cr/Au/Cr contacts. b) Scheme of the angles $\Theta_\text{M}$ and $\Theta_\text{H}$, which are the angles between the z-axis and the magnetization and external magnetic field, respectively.}
\end{center}
\end{figure}

\noindent For angular dependent spin pumping measurements the Fe(12nm)/Pt(6nm) sample has been microstructured into small stripes of 200 \textmu m$\times$ 10 \textmu m along the Fe (010) axis by means of electron beam lithography and ion beam etching. The microstructuring process did not change the spin pumping properties, since we performed vector network analyzer ferromagnetic resonance spectroscopy (VNA-FMR) measurements with the sample before and after the process. They showed that the lineshapes and the Gilbert damping did not change. Afterwards, Cr/Au/Cr contacts for the Fe/Pt stripes, a dielectric layer (SiO$_2$) and a coplanar waveguide(CPW)-like Ti/Cu/Ti antenna structure have been fabricated by electron-beam lithography, electron-beam evaporation and lacquering technique. The dielectric layer prevents crosstalk between the Cr/Au/Cr contacts and the microwave antenna. A scheme of those microstructures can be seen in Fig.~\ref{fig:oopCPW}a).

\section{VNA-FMR measurements}

\begin{figure}
\begin{center}
\includegraphics[width =0.9 \columnwidth]{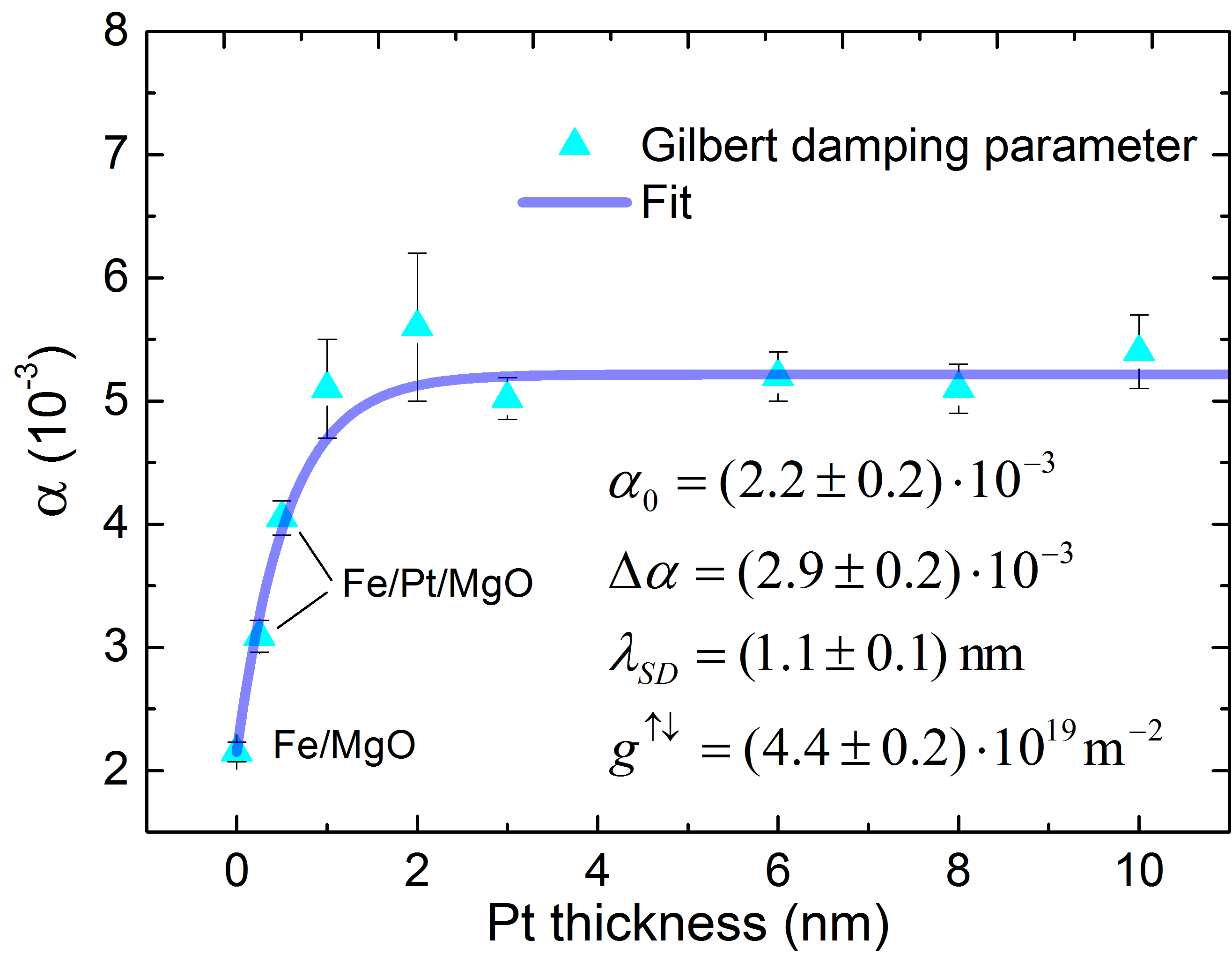}
\caption{\label{fig:alpha} Gilbert damping parameter dependence of the Pt thickness taken from ferromagnetic resonance spectroscopy measurements of the unstructured Fe/Pt samples series. Samples with Pt thicknesses below 1 nm have been additionally capped with 10 nm of MgO for corrosion protection.}
\end{center}
\end{figure}

Here, we address the VNA-FMR measurements of the unstructured Fe/Pt sample series with constant Fe thickness of 12 nm and varying Pt thicknesses. The scope of these measurements is to measure the actual spin diffusion length of Pt and the real part of the spin mixing conductance~\cite{Tserkovnyak2005}, which are critical parameters for calculating the spin Hall angle for the Fe/Pt microstructures. The Gilbert damping parameter $\mathrm{\alpha}$ is extracted by the dependence of the linewidth on the resonance frequency: 

\begin{equation}
\mu_{0} \Delta H = \mu_{0} \Delta H_{0} + \frac{ 2 \alpha f_\text{fmr}}{\gamma }
\end{equation} 
\noindent with the inhomogeneous broadening $\Delta H_{0}$ being related to the film quality. The Pt thickness dependence of $\alpha$ in Fig.~\ref{fig:alpha} can be described by the following expression~\cite{Huo2017, Shaw2012}:

\begin{equation}
\begin{split}
\alpha(t_\text{Pt}) &= \alpha_0 + \frac{ g_\text{l} \mu_{B} g^{\uparrow \downarrow} }{4 \pi M_\text{S} t_\text{Fe}} \left(1 - e^{-\frac{2 t_\text{Pt}}{\lambda_\text{SD}}} \right) \\
&= \alpha_0 + \Delta \alpha \left(1 - e^{-\frac{2 t_\text{Pt}}{\lambda_\text{SD}}} \right)~~.
\end{split}
\label{eq:lambda}
\end{equation}

\noindent with $g_\text{l}$ being the Land\'e-factor of the free electron, $\mu_{B}$ the Bohr magneton, $M_\text{S}$ the saturation magnetization, $t_\text{Fe}$ the Fe thickness and $t_\text{Pt}$ the Pt thickness. $\alpha_0$ denotes the intrinsic Gilbert damping of the Fe(12 nm) reference sample, where no spin pumping takes place. $\Delta \alpha$ is the saturation value of the Gilbert damping parameter increase due to spin pumping.

\noindent As part of Eq.~\ref{eq:lambda}, the real part of the spin mixing conductance $g^{\uparrow \downarrow}$ is the defining parameter of the transport of angular momentum through the interface~\cite{Tserkovnyak2005}.
 
The obtained value of the spin mixing conductance is then $g^{\uparrow \downarrow} = (4.4 \pm 0.2)\cdot 10^{19}$ m$^{-2}$ and stands in the same order of magnitude as for other metallic systems, like Py/Pt and CoFeB/Pt \cite{Rojas2014, Parkin2015, Ana2015}. Zhang et al. have shown with their first-principles calculations, that the real part of $g^{\uparrow \downarrow}$ can be increased up to $25\%$ in highly ordered Py/Pt bilayers by introducing interface roughness~\cite{Zhang2011}. Experimentally, this was observed in a prior publication \cite{Evangelos}, where both the interface roughness as well as $g^{\uparrow \downarrow}$ of epitaxial Fe/Pt bilayers could be enlarged by increasing the annealing temperature up to 300$^\circ$C. This optimized annealing temperature of 300$^\circ$C was also applied for the growth of the samples in this publication.

\noindent Therefore, we can use Eq.~\ref{eq:lambda} to fit the Pt thickness dependence of $\alpha$ in Fig.~\ref{fig:alpha} and extract the spin diffusion length of the epitaxial Pt $\lambda_\text{SD} = (1.1 \pm 0.1)$ nm, which is a rather small value for Pt in comparison to other authors. 
This value of $\lambda_\text{SD}$ implies that all spin currents are transmitted through the interface and are contributing to the ISHE. It has been shown that angular momentum can be lost at the interface due to spin memory loss~\cite{Rojas2014} or other interface effects may potentially also influence the Gilbert damping parameter (e.g. proximity effect~\cite{Caminale2016}). Therefore, $g^{\uparrow \downarrow}$ has to be seen as effective spin mixing conductance including all possible interface effects~\cite{Andres} 
and displays an upper limit value. Furthermore, we show through the Pt thickness dependent FMR measurements that all spin currents in Pt relevant to ISHE are located close to the Fe/Pt interface: Above a Pt thickness of about $2\lambda_\text{SD}$, all spins are flipped and additional Pt cannot contribute to the ISHE voltage. In the next section, we can also show that spin Hall angle is relatively high as well, correlating with $\lambda_\text{SD}$.

\section{Angle-resolved spin pumping measurements on microstructured Fe/Pt stripes}

Now, we are addressing the ISHE through angle-resolved spin pumping measurements:
A system, consisting of two perpendicularly aligned magnetic coil pairs, is providing an in-plane rotating external magnetic field magnitude for the spin pumping measurements. For each individual external field angle, the field is swept and the DC-voltage is measured by lock-in amplification technique. The measured loops are then fitted by Eq.~\ref{eq:ISHEfit} involving symmetric and antisymmetric Lorentzians~\cite{Ana2015}:

\begin{equation}
\begin{split}
V^z_{\text{DC}}(H) &~=~~   V^z_\text{sym} \frac {(\Delta H)^{2} }{(H-H_{\text{r}}) ^{2} + {(\Delta H)^{2} }} \\
&~+~~  V^z_\text{asym} \frac { -\Delta H(H-H_{\text{r}}) } {(H-H_{\text{r}}) ^{2} + (\Delta H)^{2} },
\end{split}
\label{eq:ISHEfit}
\end{equation}
where $\Delta H$ is the linewidth, $H_{\text{r}}$ the magnetic resonance field, $V_\text{sym}$ the symmetric and $V_\text{asym}$ the antisymmetric voltage amplitude for the case in which contacts are in z-direction (as schematically shown in Fig.~\ref{fig:oopCPW}). Each fit parameter can then be displayed in dependence of the external magnetic field angle.

\noindent To be able to interpret the angular dependence of the fit parameters, especially the voltage amplitudes, one needs to consider the microwave dynamic fields (in- and out-of-plane) and the induced microwave currents in the Fe as well as the the contact geometry. Since the Fe/Pt stripe is located between one ground line and the signal line (see Fig.~\ref{fig:oopCPW}) the out-of-plane dynamic magnetic microwave field $h_y$ is about one magnitude stronger than the in-plane field $h_x$. The other field component $h_z$ is negligible (see appendix). With the DC contacts along the z-direction, the angular dependence of the voltage amplitudes are then given by \cite{Saitoh2016, Harder2016, Keller2017}:

\begin{equation}
\begin{split}
V^{z}_{\mathrm{sym}} ~=~~ & V^{h_x}_{\mathrm{ISHE,AMR}}\sin(2\Theta_\mathrm{M})\cos(\Theta_\mathrm{M})~+\\   
& V^{h_y}_{\mathrm{AMR}}\sin(2\Theta_\mathrm{M}) ~+~ V^{h_y}_{\mathrm{AHE}} ~+\\ 
& V^{h_y}_{\mathrm{ISHE}}\sin(\Theta_\mathrm{M}) + V^{h_x}_{\mathrm{AHE}}\cos(\Theta_\mathrm{M})~~,\\								  
V^{z}_{\mathrm{asym}} =~~ & V^{h_x}_{\mathrm{AMR}}\sin(2\Theta_\mathrm{M})\cos(\Theta_\mathrm{H}) ~+~V^{h_y}_{\mathrm{AHE}} ~+\\ & V^{h_y}_{\mathrm{AMR}}\sin(2\Theta_\mathrm{M}) ~+~ V^{h_x}_{\mathrm{AHE}}\cos(\Theta_\mathrm{M})  ~~,
\end{split}
\label{eq:fit2}
\end{equation}

\noindent where $h_x$ and $h_y$ in the superscript of the voltages depict effects excited by in-plane and out-of-plane dynamic magnetic fields. In-plane excited AMR and ISHE have the same dependence ($\sin(2\Theta_\mathrm{M})\cos(\Theta_\mathrm{M})$) and can therefore not be separated ($V^{h_x}_{\mathrm{ISHE,AMR}}$ in $V^{z}_{\mathrm{sym}}$ and $V^{h_x}_{\mathrm{AMR}}$ only in $V^{z}_{\mathrm{asym}}$). In-plane ($V^{h_x}_{\mathrm{AHE}} \propto \cos(\Theta_\mathrm{M})$) and out-of-plane ($V^{h_y}_{\mathrm{AHE}} \propto \text{const.}$) excited AHE have unique shapes, but are strongly suppressed with the DC-contacts in z-direction, since microwave currents flowing in x-direction are negligible. What is expected to appear mainly, are out-of-plane excited AMR ($V^{h_y}_{\mathrm{AMR}} \propto \sin(2\Theta_\mathrm{M})$) in the symmetric and antisymmetric voltage and out-of-plane excited ISHE ($V^{h_y}_{\mathrm{ISHE}} \propto \sin(\Theta_\mathrm{M})$) in the symmetric voltage amplitude.

\begin{figure}
\begin{center}
\includegraphics[width =0.9 \columnwidth]{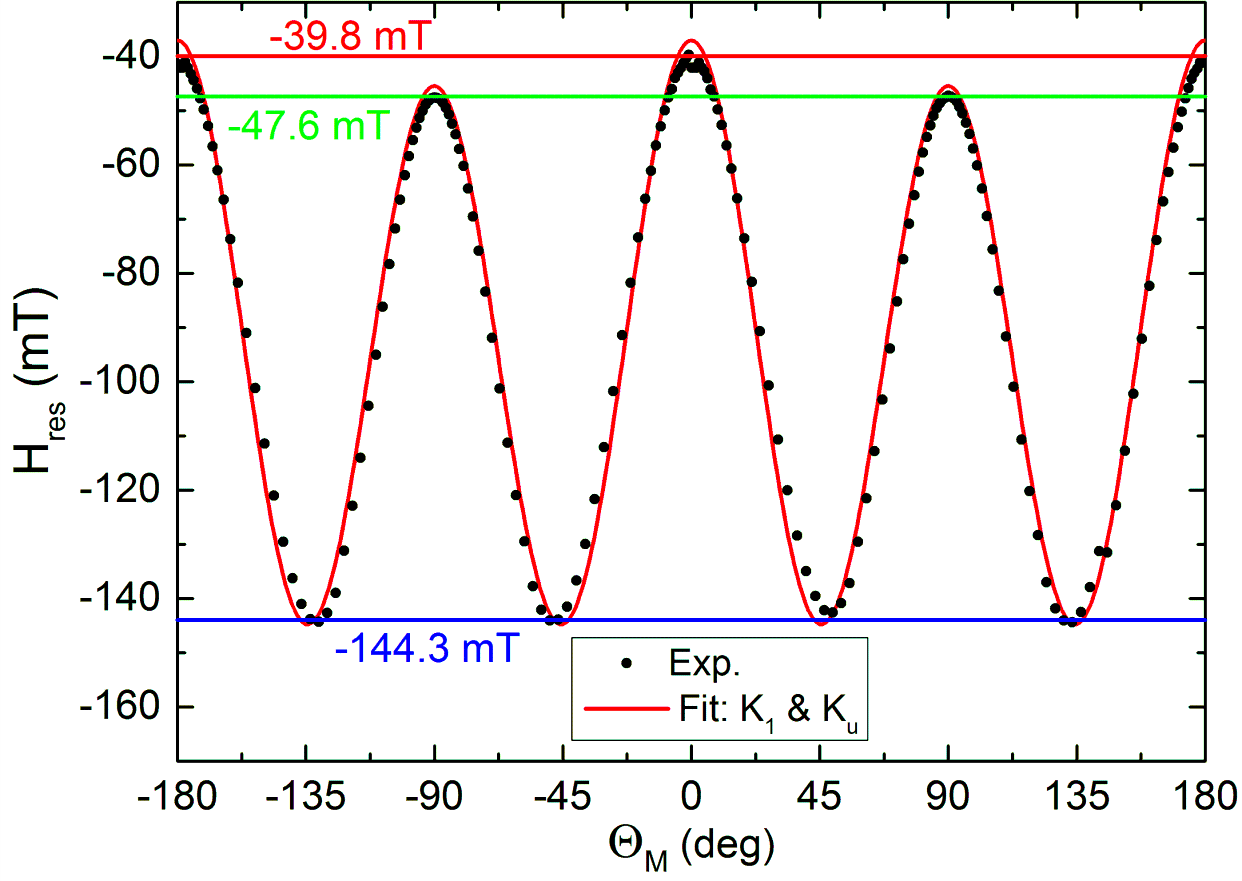}
\caption{\label{fig:fields} Resonance field against angle of magnetization of the microstructured Fe/Pt stripe obtained by the angle-resolved spin pumping measurements. The fit function of $H_\text{r}$ is according to Kittel's equation (Eq.~\ref{eq:kittel}). The maxima at -39.8 mT are the cubic and uniaxial easy axes, the maxima at -47.6 mT are the cubic easy and uniaxial hard axes and the minima at -144.3 mT are the cubic hard and uniaxial intermediate axes.}
\end{center}
\end{figure}

\noindent Since the SR effects and ISHE depend on the alignment between the magnetization and the z-axis ($\Theta_{\mathrm{M}}$), which differs from the angle between the external magnetic field and the z-axis ($\Theta_{\mathrm{H}}$), due to the anisotropy field, as schemed in Fig.~\ref{fig:oopCPW}b). The measured data sets of fit parameters in dependence of $\Theta_{\mathrm{H}}$ were readjusted to $\Theta_{\mathrm{M}}$.  Therefore, the cubic anisotropy constant $K_1$ (arising from the bcc Fe lattice) and uniaxial anisotropy constant $K_\text{u}$ (arising from shape anisotropy of the Fe/Pt stripe) are extracted from the change in resonance field $H_\text{r}$ (shown in Fig.~\ref{fig:fields}) according to:

\begin{widetext}
\begin{equation}
f_\text{r} = \mu_0 \gamma \sqrt{ \left( H_\text{r} \cos{(\Theta_\text{H}-\Theta_\text{M})}  + \left( N_x + N_x^e \right) M_\text{S}  \right)  \left( H_\text{r} \cos{(\Theta_\text{H}-\Theta_\text{M})} +\left( N_y + N_y^e \right) M_\text{S}  \right) }~~,
\label{eq:kittel}
\end{equation}
\end{widetext}

\noindent with the demagnetization factor for a thin magnetic film $N_x = 0$, $N_y = 1$ and $N_z = 0$, using effective demagnetization factors $N_x^e$ and $N_y^e$ to take the cubic and uniaxial anisotropy into account for the resonance condition (described in Eq.~\ref{eq:effdemag}) \cite{Kittel, Kohmoto2003}, and with $\gamma = 28$ GHz/T, $M_\text{S}=1712$ kA/m (measured with VNA-FMR), $K_1 = 44290 $J/m$^3$ and $K_\text{u} = 3600 $J/m$^3$.

\begin{equation}
\begin{split}
& N_y^e =  \frac{K_1}{2 \mu_0 M_\text{S}^2}(3+\cos 4\Theta_M) + \frac{2 K_u}{\mu_0 M_\text{S}^2}\cos^2 \Theta_M\\
& N_x^e =  \frac{2 K_1}{\mu_0 M_\text{S}^2}\cos 4 \Theta_M + \frac{2 K_u}{\mu_0 M_\text{S}^2}\cos 2 \Theta_M
\end{split}
\label{eq:effdemag}
\end{equation}

\begin{figure}
\begin{center}
\includegraphics[width =0.9 \columnwidth]{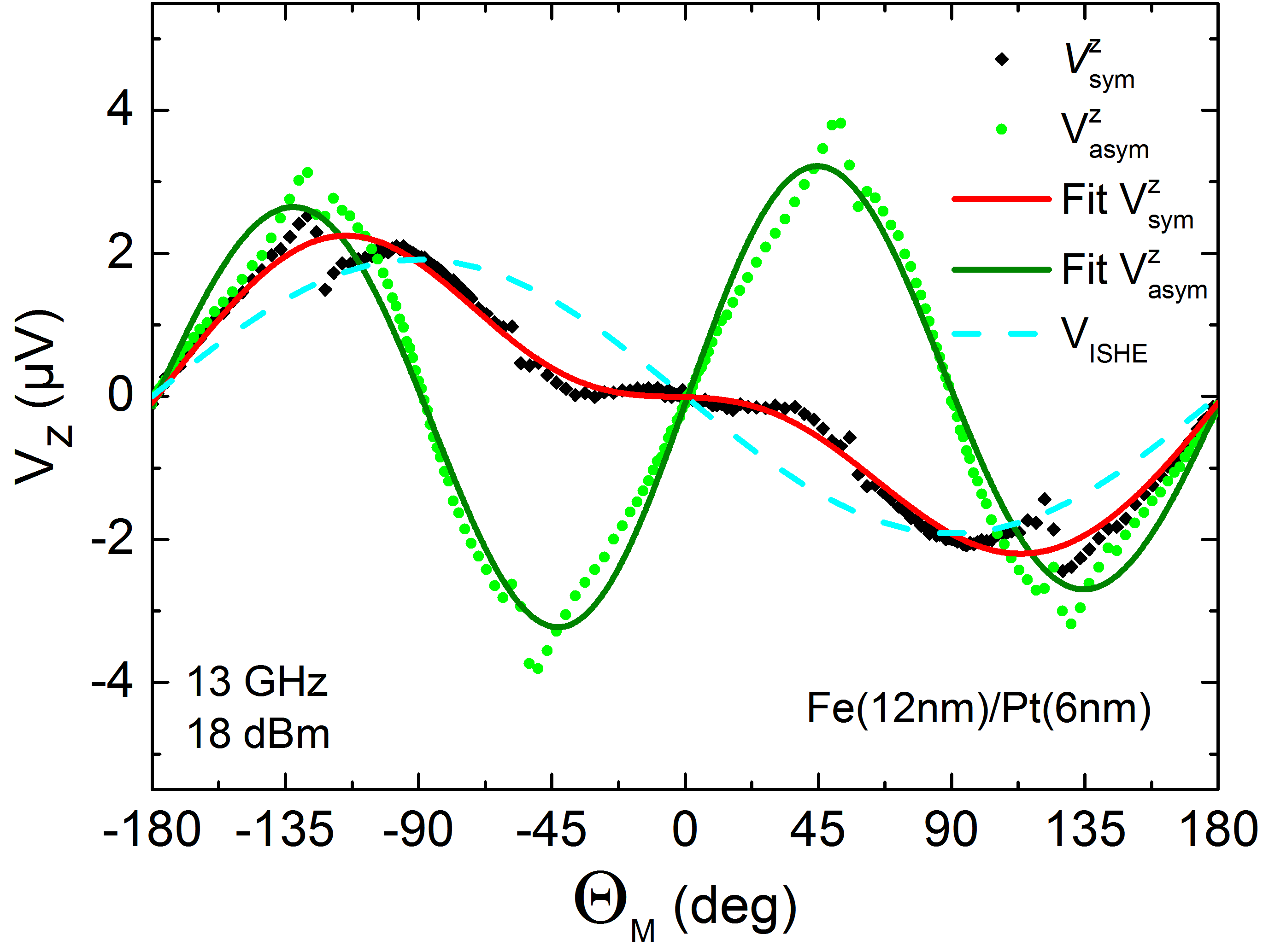}
\caption{\label{fig:voltage} Voltage amplitudes of the Fe/Pt microstructure against angle of magnetization. The out-of-plane excited ISHE, which is a component of the symmetric voltage, is shown as a blue dashed line with amplitude according to the fit. All fit functions are according to Eq.~\ref{eq:fit2}.}
\end{center}
\end{figure}

\noindent When the angle of magnetization rotates $N_y^e$ varies between 0.026 and 0.012 and $N_x^e$ varies between 0.026 and -0.024 for the given parameters. Equation~\ref{eq:kittel} is then used to calculate $\Theta_{\mathrm{M}}$ for each individual measurement point of the external field angle. The fit function parameters from Eq.~\ref{eq:ISHEfit} are then redistributed to $\Theta_{\mathrm{M}}$ and the conversion from $\Theta_\mathrm{H}$ to $\Theta_\mathrm{M}$ is performed.

\noindent In Fig.~\ref{fig:voltage} the symmetric and antisymmetric voltage amplitudes in dependence of the magnetization angle with fit functions according to Eq.~\ref{eq:fit2} are shown. The antisymmetric voltage (green dots) has almost no deviation from the expected $\sin(2\Theta_\mathrm{M})$-shape for out-of-plane excited AMR indicating low AHE (due to longitudinal contacts) and low in-plane excitation fields. The shape of the symmetric voltage (black squares) can be described by the overlapping of out-of-plane AMR and out-of-plane excited ISHE $\propto \sin(\Theta_\mathrm{H})$. The ISHE component excited by the out-of-plane dynamic magnetic field (1.92 \textmu V, component of $V_\text{sym}$ according to Eq.~\ref{eq:fit2}) is plotted as a blue dashed line. As one can see, the ISHE voltage is of similar magnitude as the AMR voltage. 
To compare all signal components, the voltages of the contributing effects are listed in Table~\ref{tab:voltages}. There, we see that effects generated by in-plane excitation fields are a magnitude smaller than effects generated by out-of-plane excitation fields and AHE contributions, in general, are negligible, since AHE with contacts in z-direction scales with $j_x$. As we show in Appendix A, the excitation by the CPW generates mainly microwave currents in z-direction and $j_x$ is negligible.
\noindent Small deviations of the data points from the fit functions can be seen for the symmetric and antisymmetric voltage amplitudes at the angles of $\pm 135^\circ$ and $\pm 45^\circ$. Around these angles double resonances appear in the measurements due to the strong magneto-crystalline anisotropy. Since our theory only account single resonances, the experimental data deviates from the fit function.

\begin{table*}
 \begin{tabular}{|c|c|c|c|c|c|}
   \hline
$V_{sym}$ or $V_{asym}$  & $V^{h_x}_{\mathrm{ISHE,AMR}}$ $($\textmu V$)$ & $V^{h_y}_{\mathrm{AMR}}$ $($\textmu V$)$  & $V^{h_x}_{\mathrm{AHE}}$ $($\textmu V$)$ & $V^{h_y}_{\mathrm{AHE}}$ $($\textmu V$)$ & $V^{h_y}_{\mathrm{ISHE}}$ $($\textmu V$)$\\
    \hline
   $V_{sym}$ & 0.20 & 0.73 & -0.05  & -0.04 & -1.92\\
    \hline
	 $V_{asym}$ & 0.39 & 2.95 & 0.02 & -0.02 & ---\\
    \hline
 \end{tabular} 
\label{tab:voltages}
\caption{ Symmetric and antisymmetric voltage amplitudes of the measured microstructures, according to Eq.~\ref{eq:fit2} and Fig.~\ref{fig:voltage}.}
\end{table*}

\section{Calculation of the spin Hall angle}

To calculate the spin Hall angle, the cone-angle of the magnetization precession $\theta$ is needed to be known. With COMSOL Multiphysics, we simulate the microwave currents inside the Fe layer and the magnetic microwave fields around it. Then, with the Polder susceptibility tensor (with relevant components $A_{xy}$ and $A_{yy}$) which transforms the magnetic excitation field $h_y$ into the dynamic components of the magnetization $m_{x'}$ and $m_y$(with the rotating coordinates $x$', $y$, $z$', where the $z$' and $x$'-axis is rotating with the magnetization), we are able to calculate the cone-angle of the FMR precession $\alpha$ with the following expression: 

\begin{equation}
P \sin^2(\theta) = A_{xy} A_{yy} h_{y}^2/M_\text{S}^2~\text{,}
\label{eq:cone}
\end{equation}

\noindent where $P$ is the correction factor for the precession ellipticity and $M_\text{S}$ is the static component of the magnetization. Further information about the COMSOL simulation, the method of calculating $h_y$ and about the Polder tensor can be found in appendix A and appendix B.\newline
Equation~\ref{eq:ISHE} describes the strength of the ISHE voltage generated by the FMR precession and the spin pumping through the Fe/Pt interface. As the magnetization precesses elliptically with the cone-angle $\theta$ at a frequency $\omega/2\pi$, a spin current proportional to $P \sin^2(\theta)$ is flowing through the Fe/Pt interface and angular momentum is transferred into the Pt, which is described by the real part of the spin mixing conductance $g^{\uparrow \downarrow}$. This spin current is then deflected by ISHE inside the Pt over the characteristic length scale $\lambda_{\text{SD}}$ with the spin-to-charge current efficiency $\Theta_{\text{SH}}$. With this, a DC voltage arises between the contact points over the stripe length $l$, which is attenuated by the electrical shunting of the metallic bilayer itself \cite{goennewein2011, Saitoh2016}. The measured ISHE voltage is then expressed by:

\begin{widetext}
\begin{equation}
V_{\text{ISHE}}^{z} = - \frac{l}{M_\text{S}^2} \sin (\Theta_\text{M})  A_{xy} A_{yy} h_{y}^2  \frac{\Theta_{\text{SH}} \lambda_{\text{SD}} g^{\uparrow \downarrow}}{\sigma_{\text{film}} t_{\text{film}}}  \left( \frac{e \omega}{2 \pi} \right) \tanh \left( \frac{t_{\text{Pt}}}{2 \lambda_{SD}} \right)~\text{,}
\label{eq:ISHE}
\end{equation}
\end{widetext}

\noindent where $l=200$ \textmu m is the length of the Fe/Pt stripe, $A_{xy}(\Theta_\text{M} = 90^\circ)= 173.7$ and $A_{yy}(\Theta_\text{M} = 90^\circ)= 35.6$ are the Polder tensor components, $\sigma_{\text{film}} = 5.4 \cdot 10^{6} $($\Omega $m)$^{-1}$ is the measured film conductivity, $t_{\text{film}}=18$ nm is the Fe/Pt film thickness, $\omega=2\pi\cdot \si{13}{\: \GHz}$ is the angular frequency of the microwave and $t_{\text{Pt}}=6$ nm is the Pt layer thickness.

\noindent With Eq.~\ref{eq:ISHE} and $h_y = 273.7~$A/m from simulation, the spin Hall angle is then calculated as $\Theta_\text{SH} = \left( 5.7 \pm 1.4\right) \%$. Therefore, the value of epitaxial Pt lies within the range of $1 \%$ up to $10 \%$ reported for mostly polycrystalline RF-sputtered Pt \cite{Hoffmann2013} and is comparable to epitaxial Pt reported by other authors \cite{Huo2017}.  
The origin of ISHE is the spin-orbit coupling of Pt, whereas theoretically there is a distinction between the intrinsic and the extrinsic mechanism behind ISHE. With the intrinsic mechanism, the spin-orbit coupling effects directly the electronic band structures. Therefore, it is independent of the electron scattering with defects, grain boundaries or phonons, contrary to the extrinsic mechanism. Intrinsic ISHE in Pt has a strong temperature dependence due to the change of resistivity. Sagasta et al. have shown~\cite{Sagasta2016} how to tune the spin Hall angle of Pt in spin valve devices by tuning the resistivity through defect density from the moderately dirty to the ultra-clean region for their Pt. They differentiate the intrinsic from the extrinsic (skew scattering) SHE contribution by the means of changing the temperature of their devices. Their ultra-clean, electron beam evaporated Pt showed a SHA of about 2 to 3 $\%$ due to their low resistivity of their Pt. This low resistance is also connected to the large Pt thickness of their devices (20 nm) compared to ours. There, the intrinsic bulk-contribution to the SHA of Pt is reduced and has the same weight as the extrinsic contribution. Although our Pt is e-beam evaporated, its low conductivity and relatively large SHA are similar to the ones of the sputtered devices. This is due to the small layer thickness of our Pt (6 nm) and furthermore due to the fact that in the very first nanometers of Pt a larger disorder in form of dislocations is expected near the Fe/Pt interface, where the lattice mismatch between Fe and Pt must be compensated (see Section 2). Despite the similarity of Sagasta’s Pt resistivity dependence of the spin Hall angle with our results, it is necessary to point out the differences: The measurements of Sagasta et al. are performed on spin valve devices with Pt, where spin currents are flowing in lateral dimensions (spin currents in spin pumping experiments are in the transversal dimension). Additionally, there the Pt has no contact with the ferromagnet. For both, the sputtered and the e-beam evaporated devices, the Pt should be polycrystalline. Therefore, the spin Hall angle of spin valve devices is not perfectly comparable to the one of spin pumping experiments. In short, the reduced conductivity of our e-beam evaporated Pt allows for the intrinsic ISHE to strongly contribute to the spin Hall angle.
Additionally to the intrinsic contribution, extrinsic mechanisms to the ISHE in our Pt could be present. The extrinsic ISHE contribution from the bulk-Pt should be largely reduced in our experiment (single-crystalline, highly ordered Pt layers) compared to polycrystalline Pt layers~\cite{Sagasta2016}. On the other side, the small spin diffusion length points to the lead that the atomically rough interface (2 to 3 monolayers) could have an extrinsic contribution (surface-assisted skew scattering) to the relatively large spin Hall angle, which was measured here at room temperature. Other authors reported similar theoretical and experimental findings: Gu et al.~\cite{Gu2010} strongly enhanced the room temperature spin Hall angle of Au films by experimentally realizing a surface-assisted spin Hall effect with Pt impurities on the Au (111) surface. In a following publication, Gu et al.~\cite{Gu2011} have shown, that this surface-assisted skew scattering mechanism strongly depends on the surface morphology of Au, on which Pt is located. Doping bulk-Au with Pt resulted in a much smaller enhancement of the spin Hall angle. Additionally, Hou et al.~investigated spin pumping and ISHE in Bi/Py bilayers and discovered, that the spin Hall angle of Bi was significantly enlarged in the vicinity of the interface and was small in the bulk-Bi~\cite{Hou2012}. And recently, Alves-Santos et al.~have greatly enlarged the spin Hall angle of Pt by inserting Ag nano-particles inside of the Pt layer, where the local spin orbit-coupling is enhanced by the Rashba effect~\cite{Alves2017}. Considering the above reported findings and the prior publication~\cite{Evangelos}, we propose the ISHE in ultra-clean epitaxial Pt is very sensitive to the conductivity of the Pt and, therefore, to the Pt layer thickness. Potentially, the interface morphology with its distinct atomic interface roughness might support ISHE by surface-assisted skew-scattering, when the conductivity of Pt is in the transition region between extrinsic and intrinsic dominated ISHE.

\section{Conclusions}

Fe/Pt bilayers have been grown stress-free with a distinct roughness of 2 to 3 monolayers at the Fe/Pt interface. With unstructured, large area Fe/Pt samples with Pt thicknesses ranging from 0 to 18 nm we determined the spin mixing conductance $(4.4 \pm 0.2)\cdot 10^{19}\: $m$^{-2}$ by VNA-FMR. By fitting the Gilbert damping dependence on the Pt thickness for the samples with very thin Pt we extracted the spin diffusion length $(1.1 \pm 0.1)$ nm, which is atypically small for Pt. With angle-resolved spin pumping measurements on a microstructured Fe/Pt stripe and with a COMSOL simulation of the microwave excitation, we obtained the spin Hall angle. Its value of $(5.7 \pm 1.4)\: \% $ is unexpectedly large for highly pure, epitaxial Pt. We explain this behavior due to the reduced conductivity of the Pt and the emerging contribution of the intrinsic mechanism.

\section*{Acknowledgments}

The Carl Zeiss Stiftung and the PPP-IKYDA 2015-DAAD bilateral German-Greek Collaboration scheme are gratefully acknowledged for financial support. We thank Tobias Fischer for support with the design of the nanostructures and thank the Team of the Nanostructuring Center Kaiserslautern for the use of their facilities and expertise.

\bibliographystyle{apsrev4-1}

\end{document}